%% file: main.tex
\documentclass[journal]{IEEEtran}

\ifCLASSINFOpdf
\else
   \usepackage[dvips]{graphicx}
\fi

\ifCLASSOPTIONcompsoc
    \usepackage[caption=false, font=normalsize, labelfont=sf, textfont=sf]{subfig}
\else
\usepackage[caption=false, font=footnotesize]{subfig}
\fi

\usepackage{url}
\usepackage{graphicx}
\usepackage{comment}

\usepackage{cite}             
\usepackage{amsmath}
\usepackage{amssymb}
\usepackage{float}

\input{miscDeclarations}

\begin{document}

\title{On the Rate-Distortion-Perception Function for Gaussian Processes}

\author{
   \IEEEauthorblockN{
        Giuseppe Serra, \IEEEmembership{Graduate Student Member, IEEE},
        Photios A. Stavrou, \IEEEmembership{Senior Member, IEEE}, 
        and Marios Kountouris, \IEEEmembership{Fellow, IEEE}}
    \thanks{The authors are with the Communication Systems Department, EURECOM, Sophia-Antipolis, France, email: \{\texttt{giuseppe.serra, fotios.stavrou, marios.kountouris\}@eurecom.fr}. M. Kountouris is also with the Andalusian Research Institute in Data Science and Computational Intelligence (DaSCI), Department of Computer Science and Artificial Intelligence, University of Granada, Spain. The work of G. Serra and M. Kountouris is supported by the European Research Council (ERC) under the European Union’s Horizon 2020 Research and Innovation Programme (Grant agreement No. 101003431). The work of P. A. Stavrou is supported in part by the SNS JU project 6G-GOALS \cite{strinati:2024} under the EU’s Horizon programme (Grant Agreement No. 101139232) and by the Huawei France-EURECOM Chair on Future Wireless Networks.}
  }

\maketitle

\begin{abstract} 
In this paper, we investigate the rate-distortion-perception function (RDPF) of a source modeled by a Gaussian Process (GP) on a measure space $\Omega$ under mean squared error (MSE) distortion and squared Wasserstein-2 perception metrics. First, we show that the optimal reconstruction process is itself a GP, characterized by a covariance operator sharing the same set of eigenvectors of the source covariance operator. Similarly to the classical rate-distortion function, this allows us to formulate the RDPF problem in terms of the Karhunen–Loève transform coefficients of the involved GPs. Leveraging the similarities with the finite-dimensional Gaussian RDPF, we formulate an analytical tight upper bound for the RDPF for GPs, which recovers the optimal solution in the \textit{``perfect realism"} regime.
Lastly, in the case where the source is a stationary GP and $\Omega$ is the interval $[0, T]$ equipped with the Lebesgue measure, we derive an upper bound on the rate and the distortion for a fixed perceptual level and $T \to \infty$ as a function of the spectral density of the source process.  
\end{abstract}


\IEEEpeerreviewmaketitle

\section{Introduction}

\IEEEPARstart{T}{he} rate-distortion-perception (RDP) trade-off, formulated simultaneously by Blau and Michaeli in \cite{blau:2019} and Matsumoto in \cite{matsumoto:2018,matsumoto:2019}, proposes a generalization of the classical rate-distortion (RD) theory \cite{shannon:59} introducing the concept of perceptual quality, i.e., the property of a sample to appear pleasing from a human perspective. This is enacted by extending the classical single-letter RD formulation, incorporating a divergence constraint between the source distribution and its estimation at the destination. The divergence constraint acts as a proxy for human perception, quantifying the satisfaction experienced when utilizing the data, as shown by its correlation with human opinion scores in \cite{mittal:2013, saad:2012}. Moreover, this divergence constraint may have multiple interpretations, such as a semantic quality metric, measuring the relevance of the reconstructed source from the observer's perspective \cite{kountouris:2020}. 

Multiple coding theorems have been developed for the RDP framework. Under the assumption of infinite common randomness between the encoder and decoder, Theis and Wagner in \cite{theis:2021} prove a coding theorem for stochastic variable-length codes in both one-shot and asymptotic regimes. Originally in the context of the output-constrained RDF, but also valid for the ``perfect realism" RDP case, Saldi {\it et. al.} \cite{saldi:2015} provide coding theorems for when only finite common randomness between encoder and decoder is available. 

Similarly to the classical RD theory, the mathematical embodiment of the RDP framework is represented by the rate-distortion-perception function (RDPF), which, as its classical counterpart, does not enjoy a general analytical solution. The absence of a general closed-form solution has prompted research into computational methods for the RDPF estimation. Toward this end, both Serra {\it et. al.} in \cite{serra:2023} and Chen {\it et. al.} in \cite{chen:2023} propose algorithms for the computation of the RDPF for general discrete sources, while \cite{serra:2024:copula} proposes a general algorithm for continuous sources for the \textit{``perfect realism"} regime, i.e., when the reconstructed process and source process are constrained to have the same statistics. 
However, despite the general complexity, certain analytical expressions have been developed for specific categories of sources. For instance, binary sources subject to Hamming distortion and total variation distance have closed-form expressions, as discussed in \cite{blau:2019}. Similarly, \cite{serra:2024:Gaussian} provides closed-form expressions for the case of scalar Gaussian sources under mean squared-error (MSE) distortion and various perceptual metrics. 

\subsection{Our Approach and Contributions}\label{subsec:contributions}
The objective of this work is the characterization of the RDPF for a source modeled by a Gaussian process (GP-RDPF), defined on a measure space $\Omega$ under MSE distortion and squared Wasserstein-2 perception metrics. To this end, we show that the optimal reconstruction of the source is itself a GP, designed such that its covariance operator shares the same set of eigenvectors with the source. The common set of eigenvalues allows the formulation of the RDPF problem as a function of the Karhunen–Loève (KL) transform coefficients of the involved GPs, similarly to the classical RDF for GPs \cite{gallager:1968, sakrison:1968:GP}. Noticing the similarities with the finite-dimensional Gaussian RDPF \cite{serra:2024:Gaussian}, we formulate an analytical upper bound to GP-RDPF able to recover the optimal solution in the \textit{``perfect realism"} regime. Lastly, focusing on the specific case where the source is a stationary GP and $\Omega$ is the interval $[0, T]$ equipped with the Lebesgue measure, we characterize a tight upper bound on the rate and distortion level for a fixed perceptual level and for $T \to \infty$ as a function of the power spectral density of the source process.

\section{Preliminaries}
\subsection{Rate-Distortion-Perception Tradeoff}
We commence by providing the definition and some properties of the RDPF for general alphabets. 

\begin{definition}(RDPF) \label{def:rdpf}
Let a source $X$ be a random variable defined on an alphabet $\mathcal{X}$ and distributed according to $P_{X} \in \mathcal{P}(\mathcal{X})$.
Then, the RDPF for $X \sim P_X$ under the distortion measure $\Delta:\mathcal{X}^2 \to \mathbb{R}^+_0$ and divergence function $d: \mathcal{P}(\mathcal{X}) \times \mathcal{P}(\mathcal{X}) \to  \mathbb{R}^+_0$ is defined as follows:
\begin{align}
    \begin{split}
        R(D,P) = & \min_{ \substack{P_{Y|X}\\ \E{\Delta(X, Y)} \le D \\
        d(P_X||P_Y) \le P}}  I(X, Y)\\
    \end{split} \label{eq:RDPF_General}
\end{align}
where the minimization is among all conditional distributions $P_{Y|X}: \mathcal{X} \to \mathcal{P}(\mathcal{X})$.
\end{definition}
We point out the following remark on Definition \ref{def:rdpf}.
  
\begin{remark}\label{remark:1}(On Definition \ref{def:rdpf}) Following \cite{blau:2019}, it can be shown that \eqref{eq:RDPF_General} has some useful {functional} properties, under mild regularity conditions.
In particular, \cite[Theorem 1]{blau:2019} shows that $R(D,P)$ is (i) monotonically non-increasing in both $D\in[D_{\min},D_{\max}]\subset[0,\infty)$ and $P\in[P_{\min},P_{\max}]\subset[0,\infty)$; (ii) convex if the divergence $d(\cdot||\cdot)$ is convex in its second argument. 
\end{remark}

\subsection{Gaussian Processes} A Gaussian Process $X$ is a collection of real random variables indexed by an index set $\set{T}$, such that for any finite subset $\set{T'} \subset \set{T}$, the collection $\{f(t)\}_{t \in \set{T'}}$ has a joint Gaussian distribution. A GP is parameterized by $m(t) = \E{X(t)}$ and $k(t,s) = \E{(X(t) - m(t))(X(s) - m(s))}$, where $m:\set{T} \to \realR$ and $k(\cdot, \cdot):\set{T}^2\to \realR$ denote the \textit{mean} and \textit{covariance} functions, respectively. It follows from its definition that $k$ is a symmetric positive semidefinite function. For a strictly positive $k$, the associated GP $f \sim \GP(m,k)$ is referred to as \textit{non-degenerate}. Furthermore, in the case $\set{T}$ is the Euclidean space $\realR^d$, a GP is said \textit{stationary} if the covariance function $k(t, s) = k (\tau)$ is a function of the difference vector $\tau = t - s$. In this case, the GP can be alternatively characterized by its power spectral density $S(f)$, i.e., the Fourier transform of its covariance function $k(\tau)$ \cite[Chapter 8]{gallager:1968}.

\subsection{Separable Hilbert spaces and $L^2$-spaces}  A Hilbert space $(\mathcal{H}, \langle \cdot,\cdot \rangle)$ is said \textit{separable} if there exists a countable orthonormal set of vectors $\mathcal{B} = \{ e_i \}_{i = 1}^{\infty}$, i.e., a Schauder basis, such that the closed linear hull of $\mathcal{B}$ spans $\mathcal{H}$ \cite{gohberg2013basic}. For a separable Hilbert space, the trace $\trace(\cdot)$ of a bounded linear operator $T$ on $\mathcal{H}$ is defined as 
$\trace(T) \triangleq \sum_{\mathcal{B}'} \langle T e_i, e_i \rangle$,
independently of the chosen basis $\mathcal{B}'$. Furthermore, a bounded operator $T$ is said \textit{trace class} if and only if $\trace[(T^*T)^{\frac{1}{2}}] < \infty$, where $T^*$ indicate the adjoint operator of $T$. 
Let $\Omega = (\set{X}, \Sigma_{\set{X}}, \mu)$ be a measure space and let $L^2(\Omega)$ be the space of $L^2$-integrable functions from $\mathcal{X}$ to $\realR$. Equipping $L^2(\Omega)$ with the inner product $\langle f,g \rangle  = \int_{\Omega} f(x) g(x) \mu(dx)$, for $f,g \in L^2(\Omega)$, it becomes an Hilbert space\footnotemark. Throughout this paper, we assume that the underlying measure space $\set{X}$ is such that $(L^2(\Omega), \langle \cdot,\cdot \rangle )$ is separable. 

\footnotetext{Formally, $\langle \cdot, \cdot \rangle$ is a semi-inner product. However, introducing the equivalence classes of functions that differ only on $\mu$-negligible sets, i.e. $f = g \iff f - g = 0 ~\mu-a.e.$ , $\langle \cdot, \cdot \rangle$ becomes an inner product.}

\subsection{Covariance Operators and Mercer's Theorem}
Given a covariance function $k \in L^2(\Omega \times \Omega)$, we can define the associated Hilbert–Schmidt (HS) integral operator $K:L^2(\Omega) \to L^2(\Omega)$ as
\begin{align*}
    [K\phi](t) = \int_{\Omega} k(x,s) \phi(x) \mu(dx) \qquad t \in \set{X}.
\end{align*}
Then, $K$ is a self-adjoint, compact, positive, and trace class operator, and the space of such covariance operators is a convex space. Moreover, since the mapping $k \to K$ is an isometric isomorphism from $L^2(\Omega \times \Omega)$ to the space of Hilbert-Schmidt operators on $L^2(\Omega)$, we use interchangeably the notations $\GP(m,k)$ and $\GP(m, K)$. \\
The constructed HS operator $K$ also satisfies the conditions of Mercer's Theorem \cite{Steinwart:2012}. Let $\{\phi_i\}_{i = 1}^{\infty}$ and $\{\lambda_i\}_{i = 1}^{\infty}$  be the sets of eigenfunctions and associated eigenvalues of $K$. Then, the covariance function $k$ can be represented by the expansion
\begin{align*}
    k(t,s) = \sum_{i = 1}^{\infty} \lambda_i \phi_i(t) \phi_i(s) \qquad (t,s) \in \set{X}^2,
\end{align*}
with absolute and uniform convergence on $\Omega \times \Omega$, implying that $K$ can be expressed as
\begin{align*}
    [K\psi](t) = \sum_{i = 1}^{\infty} \lambda_i \langle \psi, \phi_i \rangle \phi_i(t) \qquad t \in \set{X},
\end{align*}
i.e., $K$ is a diagonizable operator. Furthermore, a zero-mean stochastic process $X$ can be represented as
\begin{align*}
    X(t) = \sum_{i = 1}^\infty X_i \phi_i(t) \qquad t \in \set{X}
\end{align*}
where $X_i = \langle X, \phi_i \rangle$ is a random variable with $\E{X_i} = 0$ and $\E{X_iX_j} = \lambda_i \delta_{i,j}$. Additionally, if $X \sim \GP(0,K)$, then $X_i$ is Gaussian distributed and, consequently, $\forall (i,j) ~ X_i \perp X_j$. Therefore, given a suitable base of $L^2(\Omega)$, the GP $X$ can be represented by a countable set $\{X_i\}_{i = 1}^{\infty}$ of independent Gaussians. This representation is usually referred to as KL transform \cite{berlinet:2011:reproducing} and we refer to $\{X_i\}_{i = 1}^{\infty}$ as the KL coefficients of the process $X$.

\subsection{Squared Wasserstein-2 distance for GP} 
Squared Wasserstein-2 distance was originally introduced in \cite{gelbrich:1990:W2formula} as a specific instance of the optimal transport problem (see, e.g., \cite[Chapter 7]{villani:2021}). In particular, squared Wasserstein-2 distance is defined as follows
\begin{align}
    \W2(P_X, P_{Y}) {\triangleq} \min_{\Pi(P_X, P_{Y})} \E{||X - Y||^2}
\end{align}
where $\Pi(P_X, P_{Y})$ is the set of all joint distributions {$P_{X,Y}$} with  marginals $P_X$ and $P_{Y}$.
Following \cite[Definition 1]{mallasto:2017}, the Wasserstein-2 distance can be extended to GPs. Let $X \sim \GP(m_X, k_X)$ and $Y \sim \GP(m_Y, k_Y)$ with $m_X,m_Y \in L^2(\Omega)$ and $k_X, k_Y \in L^2(\Omega \times \Omega)$, then the squared Wasserstein-2 distance between GPs $X$ and $Y$ is given by
\begin{align*}
    \W2(X,Y) &= d^2(m_X,m_Y) \\
    &\quad+ \trace \left[K_X + K_Y - 2\left(K_X^{\frac{1}{2}}K_YK_X^{\frac{1}{2}}\right)^{\frac{1}{2}}\right]
\end{align*}
where $d^2(\cdot,\cdot)$ is the canonical distance in $L^2(\Omega)$ and $K_X, K_Y$ are the HS operators associated with $k_X$ and $k_Y$, respectively.

\section{Main Results}
We start this section by providing a formal characterization of the RDPF problem for sources modeled as GPs.

\begin{theorem}\label{th:GP-RDPF}
(GP-RDPF) Let $D \ge 0$, $P \ge 0$, and $X \sim \GP(0, K_X)$ be a source modeled by a GP. Then, the associated RDPF under MSE distortion and Wasserstein-2 divergence is achieved by a reconstruction $Y \sim \GP(0, K_Y)$ (i.e., is itself a GP), such that $K_X$ and $K_Y$ share the same set of eigenvectors. Additionally, the RDPF can be expressed as
\begin{align}
    \begin{split}
        R_{GP}(D,P) = & \min_{\{P_{Y_i|X_i}\}_{i = 1}^{\infty}} \sum_{i = 1}^{\infty} I(X_i;Y_i) \\
        \textrm{s.t.}   & \quad \sum_{i = 1}^{\infty} \E{||X_i - Y_i||^2} \le D\\
        & \quad \sum_{i = 1}^{\infty} \W2(X_i,Y_i) \le P 
    \end{split} \label{eq:RDPF_GP}
\end{align}
where $\{X_i\}_{i = 1}^{\infty}$ and $\{Y_i\}_{i = 1}^{\infty}$ are the KL coefficients of the GP $X$ (source) and the GP $Y$ (reconstruction), respectively. 
    
\end{theorem} 
\begin{IEEEproof} Before we delve into the technicalities of the proof, we give some useful notation. We denote with $\{ \phi_i\}_{i = 0}^{+\infty}$ and $\{\lambda_i\}_{i = 0}^{+\infty}$ the set of eigenvectors and eigenvalues of $K_X$. Moreover, let $\{ \eta_i \}_{i = 0}^{+\infty} \subset L^2(\Omega)$ be any countable orthonormal set of eigenvectors. Then, we can construct the HS operator $K_Y$ and the associated process $Y$ as
    \begin{align}
        [K_Y \psi] = \sum_{i = 1}^{\infty} \nu_i \langle \psi, \eta_i \rangle \eta_i \qquad Y = \sum_{i = 1}^{\infty} Y_i \eta_i,    
    \end{align}
with $\E{Y_i} = 0$ and $\E{Y_iY_j} = \nu_i \delta_{i,j}$. As a result, the mutual information between the processes $X$ and $Y$ can be expressed as $I\left(\{X_i\}_{i = 1}^{\infty}; \{Y_i\}_{i = 1}^{\infty}\right)$. Note that, until now, we did not assume that $\{Y_i\}_{i = 0}^{\infty}$ are necessarily Gaussian distributed.
    
We now show that the assumption of $\{\eta_i\}_{i = 1}^{\infty} = \{\phi_i\}_{i = 1}^{\infty}$ and $\{Y_i\}_{i = 1}^{\infty}$ Gaussian distributed results optimal. Leveraging the equivalence between GP and non-degenerate Gaussian measures, \cite[Proposition 2.4]{CAlbertos:1996} allows to lower bound the $\W2(\cdot,\cdot)$ perception as follows
    \begin{align*}
        \W2(f_X, f_Y) & \stackrel{(a)}{\ge} \sum_{i = 1}^{\infty} \W2(X_i,Y_i)\\
        & \stackrel{(b)}{\ge} |\mu_{X_i} - \mu_{Y_i}|^2 + \sum_{i = 1}^{\infty} (\sqrt{\lambda_i} - \sqrt{\eta_i})^2
    \end{align*}
where (a) and (b) hold with equality iff $\{\eta_i\}_{i = 1}^{\infty} = \{\phi_i\}_{i = 1}^{\infty}$ and $\{X_i\}_{i = 1}^{\infty}$ and $\{Y_i\}_{i = 1}^{\infty}$ are Gaussian distributed, respectively. 
The optimality of the considered assumptions with respect to the mutual information and the MSE distortion stems from their proven optimality in the classical RDF \cite{sakrison:1968:GP}. Consequently, the MSE of the two processes can be expressed as
    \begin{align*}
        \E{||X - Y||^2} &= \E{\left|\left|\sum_{i = 0}^{\infty} (X_i - Y_i) \phi_i \right|\right|^2} = \sum_{i = 0}^{\infty} \E{|X_i - Y_i|^2}.
    \end{align*}
Furthermore, since $\{Y_i\}_{i = 0}^{\infty}$ is a set of uncorrelated Gaussian random variables, and therefore independent, the mutual information $I\left(\{X_i\}_{i = 1}^{\infty}; \{Y_i\}_{i = 1}^{\infty}\right)$ becomes
    \begin{align*}
        I\left(\{X_i\}_{i = 1}^{\infty}; \{Y_i\}_{i = 1}^{\infty}\right) = \sum_{i = 1}^{\infty} I(X_i,Y_i).
    \end{align*}
This concludes the proof.
\end{IEEEproof}
Theorem \ref{th:GP-RDPF} provides a structural characterization of the RDPF problem for GPs and serves as an optimization problem on the set of joint random variables $\{(X_i, Y_i)\}_{i =1}^{\infty}$ defining the statistics of the source and reconstruction GPs. This process can be seen as selecting the proper set of orthonormal vectors, i.e., the eigenvectors of $K_X$, so that the problem "diagonalizes", similarly to the finite-dimensional Gaussian RDPF \cite{serra:2024:Gaussian}. \\
The following corollary further simplifies \eqref{eq:RDPF_GP} leveraging the knowledge of the analytic solution of the scalar Gaussian RDPF under MSE distortion and $\W2(\cdot,\cdot)$ perception.

\begin{corollary} The optimization problem in \eqref{eq:RDPF_GP} can be cast as follows
\begin{align}
    \begin{split}
        R_{GP}(D,P) = & \min_{
        \substack{\{(D_i,P_i)\}_{i = 1}^{\infty}\\
        \sum_{i = 1}^{\infty} D_i \le D \\ \sum_{i = 1}^{\infty} P_i \le P
        }} \sum_{i = 1}^{\infty} R_{X_i}(D_i,P_i) \\
    \end{split} \label{eq:RDPF_GP_Alt}
\end{align}
where $R_{X_i}(\cdot, \cdot)$ is the RDPF under MSE distortion and squared Wasserstein-2 perception for the Gaussian $X_i \sim \mathcal{N}(0, \lambda_i)$. 
    
\end{corollary}
\begin{IEEEproof}
Introducing the additional constraints $\E{|X_i - Y_i|^2} \le D_i$ and $\W2(X_i, Y_i) \le P_i$, we can express \eqref{eq:RDPF_GP} as
    \begin{align*}
        R_{GP}(D,P) = & \min_{\substack{\{(D_i,P_i)\}_{i = 1}^{\infty}\\
        \sum_{i=1}^{\infty} D_i \le D\\
        \sum_{i=1}^{\infty} P_i \le P}} \sum_{i = 1}^{\infty} \min_{\substack{P_{Y_i|X_i}\\ \E{|X_i - Y_i|^2} \le D_i \\ 
        \W2(X_i,Y_i) \le P_i}} I(X_i, Y_i)
    \end{align*}
where each element in the summation can be recognized to be the definition of the RDPF $R_{X_i}(D_i, P_i)$ for a Gaussian source $X_i$. This concludes the proof.
\end{IEEEproof}

Serra \textit{et. al.} \cite{serra:2024:Gaussian} analyze the finite-dimensional version of \eqref{eq:RDPF_GP_Alt}, designing a solving algorithm based on alternating minimization for the computation of the optimal allocations $\{(D_i, P_i)\}_{i = 1}^{\infty}$. Their solution leverages an alternating minimization scheme, where $\{D_i\}_{i = 1}^{\infty}$ and $\{P_i\}_{i = 1}^{\infty}$ are alternatively optimized while fixing the other set of variables. Alas, this computational approach cannot be implemented in \eqref{eq:RDPF_GP_Alt} due to the cardinality of the set of optimization variables. Nonetheless, fixing the perceptual levels $\{P_i\}_{i = 1}^{\infty}$ allows us to characterize the optimal distortion allocation and associated per-dimension rates, as shown in the following theorem.

\begin{theorem} \label{th: optimal_distortion_alloc}
Let the perception allocations $\{P_i\}_{i = 1}^{\infty}$, such that $\sum_{i = 1}^{\infty} P_i \le P$, be given. Then, the associated optimal distortion allocations $\{\tilde{D}_i\}_{i = 1}^{\infty}$ and the per-dimension rates $\{\tilde{R}_{X_i}\}_{i = 1}^{\infty}$ are
    \begin{align*}
        \tilde{D}_i = 
        \begin{cases} \min \left\{\gamma, \lambda_i \right\} \qquad  \textit{if} ~ \sqrt{P_i} \ge \sqrt{\lambda_i} - \sqrt{\lambda_i - \min\{\lambda_i, \gamma\}} \\\\
        P_i + 2\sqrt{\lambda_i}\left(\sqrt{\lambda_i} - \sqrt{P_i}\right) \\ 
        \quad  + \gamma - \sqrt{ 4\lambda_i\left(\sqrt{\lambda_i} - \sqrt{P_i}\right)^2 + \gamma^2}  \qquad ~~~~\textit{otherwise.}
        \end{cases}
       \end{align*}
    \begin{align*}
        \tilde{R}_{X_i} = 
        \begin{cases} \frac{1}{2} \log^+ \left(\frac{\lambda_i}{\gamma}\right) ~ ~\textit{if} ~ ~ \sqrt{P_i} \ge \sqrt{\lambda_i} - \sqrt{\lambda_i - \min\{\lambda_i, \gamma\}} \\\\
        \frac{1}{2} \log \left( \frac{2 \lambda_i (\sqrt{\lambda_i} - \sqrt{P_i})^2}{\gamma \left(\sqrt{4\lambda_i(\sqrt{\lambda_i}-\sqrt{P_i})^2 + \gamma^2} - \gamma \right)} \right) \qquad ~ \textit{otherwise.}
        \end{cases}
    \end{align*}
    where $\log^+(x) = \max\{x, 0\}$ and $\gamma \ge 0$ is chosen such that $\sum_{i = 1}^{\infty} \tilde{D}_i \le D$.
\end{theorem}
\begin{IEEEproof}
    The proof of the optimal distortion allocation $\{D_i\}$ follows as a limit case of \cite[Theorem 7]{serra:2024:Gaussian} for the finite-dimensional case. The additional assumption $\sum_{i = 1}^{\infty} P_i \le P$ is required to ensure that $\sum_{i = 0}^{\infty} \tilde{D}_i < \infty$ independently of $\gamma$, since it can be shown that
    \begin{align*}
        \sum_{i = 1}^{\infty} \tilde{D}_i < \sum_{i = 1}^{\infty} P_i + 2\sum_{i = 1}^{\infty} \lambda_i \stackrel{(a)}{<} +\infty
    \end{align*}
    where (a) derives from $\{\lambda_i\}_{i = 1}^{\infty}$ being the eigenvalues of the trace-class operator $K_X$. The associate per-dimension rates $\{R_{X_i}\}_{i = 1}^{\infty}$ results from the scalar Gaussian RDPF \cite{serra:2024:Gaussian} evaluated at $(\tilde{D_i}, P_i)$. This concludes the proof.
\end{IEEEproof} 
We stress the following technical remark for Theorem \ref{th: optimal_distortion_alloc}.
\begin{remark} \label{remark: comparison with WF}
Fixed the perceptual levels $\{P_i\}_{i=1}^{\infty}$, the parametric solutions recovered in Theorem \ref{th: optimal_distortion_alloc} can be interpreted as an extension of the classical water-filling solution for the GP-RDF. In fact, we remark that the first branch of the function $\tilde{D_i}$ and $\tilde{R}_{X_i}$ recovers the classical RDF solutions \cite{sakrison:1968:GP}.
Conversely, the second branch of the functions can be interpreted as an adaptive water-level solution; $\tilde{D}_i$, through the dependence on the $i^{th}$ dimension second moment $\lambda_i$, gets adapted to each dimension, thus guaranteeing that all source components are present in the reconstructed signal.
\end{remark}

It should be noted that the analytical characterization of the optimal perception levels $\{P^*_i\}_{i = 1 }^{\infty}$ remains an open problem, even in the finite dimension setting. However, for any convergent series $\{\tilde{P}_i\}_{i = 1}^{\infty}$, Theorem \ref{th: optimal_distortion_alloc} characterizes the associated minimizing distortion allocation $\{\tilde{D}_i\}_{i = 1}^{\infty}$ and the per-dimension rates $\{\tilde{R}_{X_i}\}_{i = 1}^{\infty}$, which identify an analytical upper bound to the GP-RDPF, i.e.,
\begin{align*}
    R_{GP}(D,P) = \min_{\{D_i,P_i\}_{i = 1}^{\infty}}\sum_{i = 1}^{\infty} R_{X_i}(D_i, P_i) \le \sum_{i = 1}^{\infty} \tilde{R}_{X_i}.
\end{align*}
Nevertheless, the numerical results for the finite-dimensional Gaussian RDPF from \cite[Alg. 1]{serra:2024:Gaussian} hint at a relative proportionality in the perceptual levels assignment. Based on this observation, we propose a heuristic perceptual levels allocation proportional to the second moments of $\{X_i\}_{i = 1}^{\infty}$, i.e., 
\begin{align}
    \tilde{P}_i = \alpha \lambda_i \quad \Longrightarrow \quad \alpha\sum_{i = 1}^{\infty} \lambda_i = P \label{eq:heuristic_P_allocation}
\end{align}
where $\alpha$ acts as a proportionality constant to enforce the desired perception level $P$.
The advantage of this allocation is that it recovers the optimal solution for $P \to 0$, i.e., for all $i = 1,2,\ldots$ $P_i \to 0$, while still providing an excellent bound in the general case.

\subsection{GP-RDPF for Stationary Sources}
We devote this section to characterizing the particular case of GP-RDPF for stationary processes. To this end, we consider $\Omega = (\set{X},\Sigma_{\set{X}},\mu)$ such that $\set{X} = [0,T]$, $\Sigma_{\set{X}}$ is the $\sigma$-algebra of all Lebesgue measurable subsets $\mathcal{U} \subseteq [0,T]$, and $\mu$ is the Lebesgue measure. Similarly to the classical RDF \cite{gallager:1968, sakrison:1968:GP}, in this setting, we formulate the RDPF considering normalized versions of main quantities of interest, i.e.,
\begin{align*}
    R = \frac{1}{T} \sum_{i = 1}^{\infty} R_{X_i}(D_i,P_i), \quad D \ge \frac{1}{T} \sum_{i = 1}^{\infty} D_i, \quad P \ge \frac{1}{T} \sum_{i = 1}^{\infty} P_i. 
\end{align*}
Under the assumption that the source $X$ is a stationary GP, this normalization allows us to extend the results of Theorem \ref{th: optimal_distortion_alloc} to the case where $T \to \infty$, as shown in the following theorem.
\begin{theorem} \label{th:StationaryRDPF} 
Let $X \sim \GP(0,k_X)$ be a stationary process defined on $[0,T] \subset \realR$ and let $S_X$ be the associated power spectral density. Then, for $T \to \infty$, the GP-RDPF is upper bounded by the parametric curve $(\tilde{R},\tilde{D},\tilde{P})$ parameterized by $\gamma > 0$ and $0 \le \alpha \le 1$ as  
\begin{align} 
    \tilde{R} &= \frac{1}{2} \int_{\mathcal{S}_{RDP}} \log\left( \frac{2 S^2_X(f)(1-\sqrt{\alpha})^2 }{ \gamma (\sqrt{\gamma^2 + 4 S^2_X(f)(1-\sqrt{\alpha})^2} - \gamma)} \right) df \nonumber\\
    & \quad + \frac{1}{2}\int_{\mathcal{S}_{RD}} \log^+\left(\frac{S_X(f)}{\gamma}\right) df \label{eq: Rate_Stationary}\\
    \tilde{D}   &= \int_{\mathcal{S}_{RDP}} \hat{D}(S_X(f)) df + \int_{\mathcal{S}_{RD}} \min\{\gamma, S_X(f)\} df \label{eq: Distortion_Stationary}\\
    & ~~~\hat{D}(x) = \gamma + x(1+(1-\sqrt{\alpha})^2) - \sqrt{4x^2(1-\sqrt{\alpha})^2 + \gamma^2} \nonumber\\
    \tilde{P} &= \alpha \int S_X(f) df
\end{align}
where the sets $\mathcal{S}_{RD}$ and $\mathcal{S}_{RDP}$ are defined as
\begin{align*} 
\mathcal{S}_{RD} &\triangleq \{f\in \realR: S_X(f)  \ge \gamma + (1 - \sqrt{\alpha})^2 S_X(f) \}\\
\mathcal{S}_{RDP} &\triangleq\realR / \mathcal{S}_{RD}.
\end{align*}

\end{theorem}
\begin{IEEEproof} By considering the proportional perceptual level allocation described in \eqref{eq:heuristic_P_allocation}, the proof of the theorem follows as a direct application of \cite[Lemma 8.5.3]{gallager:1968} to the results of Theorem \ref{th: optimal_distortion_alloc}.
\end{IEEEproof}

\section{Numerical Examples}
\begin{figure}[h!]
    \centering
    \includegraphics[width = \linewidth]{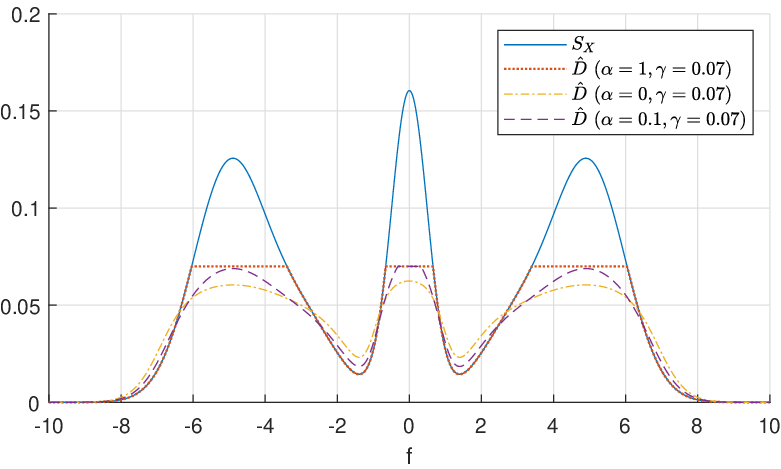}
    \caption{Source Power Spectrum $S_X(f)$ vs. per-frequency distortion $\hat{D}(S_X(f))$ for $\gamma = 0.7$ and varying $\alpha$.}
    \label{fig:SpectrumVsDistortionComparison}
\end{figure}

In this section, we provide a numerical example to better illustrate the results of Theorem \ref{th:StationaryRDPF}. Let the source $X$ be modeled as a stationary GP with power spectral density $S_X$ as reported in Fig. 
\ref{fig:SpectrumVsDistortionComparison}.

We first investigate the per-frequency distortion $\hat{D}(S_X(f))$, defined as the integral argument of \eqref{eq: Distortion_Stationary}, for fixed $\gamma$ and varying $\alpha \in [0,1]$. For $\alpha = 1$, i.e., no perceptual constraint enforced, the distortion allocation follows the classical RDF water-filling solution \cite{gallager:1968}. For lower values of $\alpha$, i.e., stricter perceptual requirements, the distortion allocation adapts to the structure of the source power spectrum $S_X(f)$, extending the observations in Remark \ref{remark: comparison with WF} to the stationary sources.
\par
\begin{figure}[t!]
    \centering
    \subfloat[\label{1a}]{   \includegraphics[width=\linewidth]{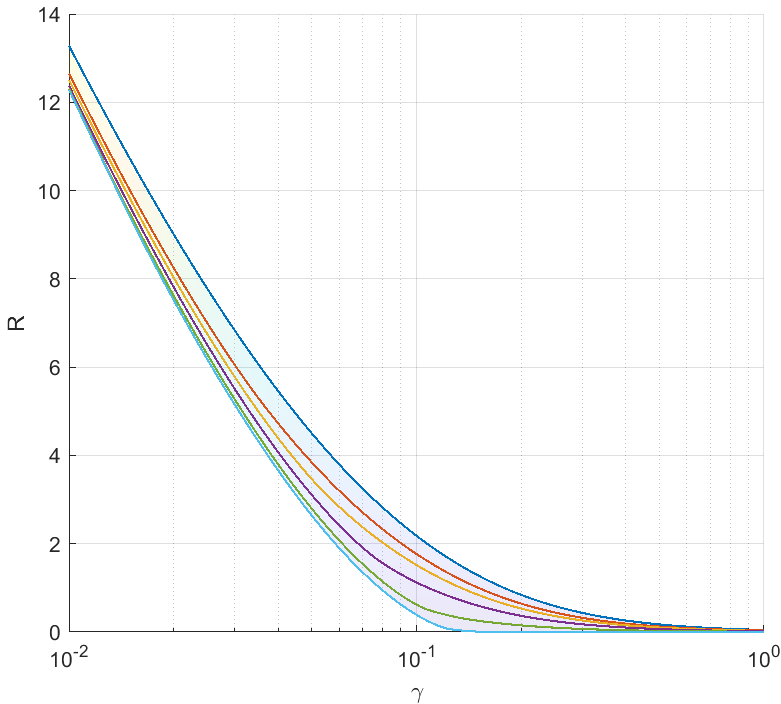}
    }\\
    \subfloat[\label{1b}]{
    \includegraphics[width=\linewidth]{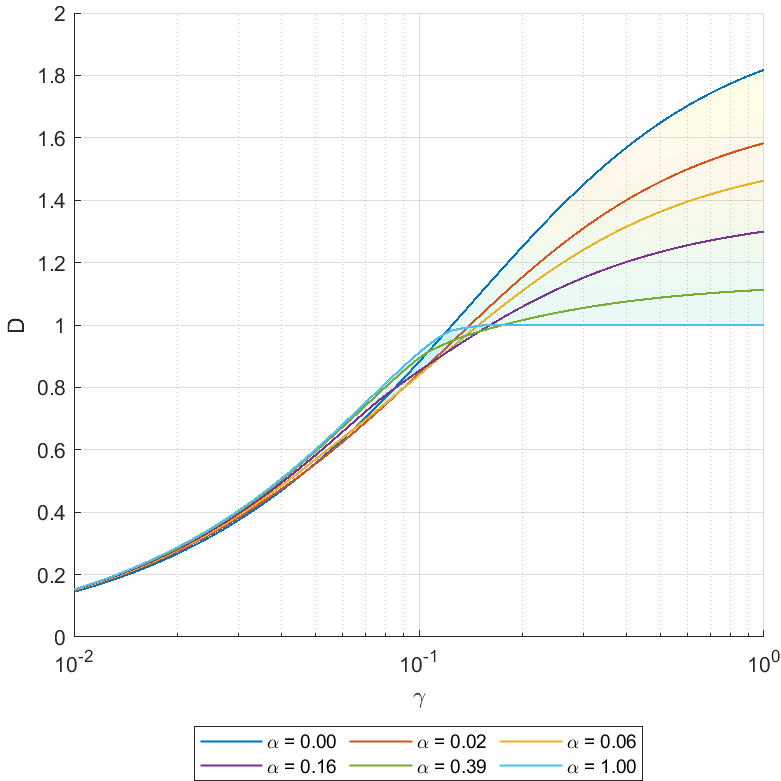}
    }
    \caption{(a) Rate $\tilde{R}$ and (b) distortion $\tilde{D}$ curves parameterized by $(\alpha, \gamma)$.}
        \label{fig:RDparametrized}
\end{figure}

 Fig. \ref{fig:RDparametrized} shows the numerically estimation curves of the rate $\tilde{R}$ (Fig. \ref{1a}) and distortion $\tilde{D}$ (Fig. \ref{1b}) defined in \eqref{eq: Rate_Stationary} and \eqref{eq: Distortion_Stationary}, respectively. The curves are computed considering parameters $\gamma \in [0.01,1]$ and $\alpha \in [0,1]$, showing similar properties to the Gaussian RPDF in the finite dimensional setting.

\section{Conclusion}

In this paper, we characterized the RDPF of a source modeled as a GP considering MSE distortion and squared Wasserstein-2 perceptual metrics. We first provided a general characterization for a non-stationary source in terms of its Karhunen–Loève representation, leveraging the properties of the distortion and perceptual constraints. We then formulated an analytical upper bound, able to capture the exact RDPF in the ``perfect realism" regime. Lastly, we extended the obtained closed-form result to the case of a stationary GP on the real line, expressing the bound as a function of the source spectral power density.

\bibliographystyle{IEEEtran}
\bibliography{strings, biblio}

\end{document}

%% file: miscDeclarations.tex
\usepackage{amsmath}
\usepackage{amssymb}
\usepackage{xcolor}

\newtheorem{theorem}{\bfseries Theorem}

\newtheorem{definition}{\bfseries Definition}
\newtheorem{corollary}{\bfseries Corollary}

\newtheorem{remark}{\bfseries Remark}

\DeclareMathOperator{\trace}{Tr}

\DeclareMathOperator{\W2}{W_2^2}

\newcommand{\realR}{\mathbb{R}}

\newcommand{\GP}{\mathcal{GP}}

\newcommand{\E}[2][]{ 
    \mathbb{E}_{#1} \left[ #2 \right]
}

\newcommand{\set}[1]{\mathcal{#1}}